\documentstyle[aps,preprint,prc,12pt]{revtex}
\newcommand{\be}{\begin{equation}}
\newcommand{\ee}{\end{equation}}

     \makeatletter
     \def\thefigure{\@arabic\c@figure}\def\fps@figure{tbp}
     \def\ftype@figure{1}\def\ext@figure{lof}
     \def\fnum@figure{\protect\footnotesize Fig.\ \thefigure}
     \def\thetable{\@arabic\c@table}
     \def\fps@table{tbp}\def\ftype@table{2}\def\ext@table{lot}
     \def\fnum@table{\protect\footnotesize Table \thetable}
     \def\@listI{\leftmargin\leftmargini\parsep=0pt\itemsep=0pt}
     \def\thebibliography#1{\section{References}\vspace*{-10pt}\list
    {[\arabic{enumi}]}{\settowidth\labelwidth{[#1]}\leftmargin\labelwidth
     \advance\leftmargin\labelsep
      \usecounter{enumi}}
     \def\newblock{\hskip .11em plus .33em minus .07em}
      \sloppy\clubpenalty4000\widowpenalty4000
     \sfcode`\.=1000\relax}
     
     \def\@nomath#1{\ifmmode \fi}
     \def\mmycite{\@ifnextchar [{\@tempswatrue\@mmycitex}
      {\@tempswafalse\@mmycitex[]}}
      \def\@mmycitex[#1]#2{\if@filesw\immediate%
     \write\@auxout{\string\citation{#2}}\fi
       \def\@citea{}\@mmycite{\@for\@citeb:=#2\do
      {\@citea\def\@citea{,}\@ifundefined
        {b@\@citeb}{{\bf ?}\@warning
     {Citation `\@citeb' on page \thepage \space undefined}}%
     \hbox{\csname b@\@citeb\endcsname}}}{#1}}
     \def\@mmycite#1#2{{{\scriptsize#1}\if@tempswa , #2\fi}}
     \def\mycite#1{$^{\protect\mmycite{#1}}$}
      \def\@cite#1#2{{#1\if@tempswa , #2\fi}}
      \def\thesection {\arabic{section}}
      \def\section#1{\addtocounter{section}{1}\setcounter{subsection}{0}
       \bigskip\medskip{\noindent\bf\thesection.\ #1}
        \medskip}
     \def\thesubsection {\arabic{section}.\arabic{subsection}}
        \def\subsection#1{\addtocounter{subsection}{1}
           \medskip{\noindent\thesubsection.\ #1}
           \medskip}
       \makeatother
\begin{document}

\title{\bf DYNAMICS OF FRAGMENT FORMATION
 IN THE NUCLEAR SPINODAL REGION}

\author {M.\, Baldo, G.\,F. Burgio and A.\, Rapisarda}

\address  {\it Istituto Nazionale
di Fisica Nucleare, Sezione di Catania}
\address  {\it   and Dipartimento
 di Fisica, Universit\'a di Catania }
\address  {\it  Corso Italia 57, I-95129 Catania, Italy }

\date{ July 12, 1994}
\maketitle

\begin{abstract}
The Vlasov-Nordheim equation is solved numerically on a lattice for
nuclear matter in two dimensions. We discuss the reliability of the
model at normal density and then study the response of the system to
small perturbations. We find deterministic chaos inside the spinodal
zone  where fragment formation occurs. We discuss in detail the
dynamical features of this phenomenon in order to clarify the
mechanisms leading to nuclear disassembly in heavy--ion collisions.
\end{abstract}

\bigskip
Pacs: 25.70.Pq, 24.60.Lz, 21.65.+f

\newpage
Many recent experiments\mycite{hir94} in heavy--ion collisions at
energies above 50 MeV/A  have shown a very rapid break-up of the hot
composite system  formed in the reaction into several big fragments
with $Z \ge 3$.  This disassembly of the nuclear system, which can
include a substantial fraction of the  colliding masses, falls under
the generic name of  "nuclear multifragmentation".  In parallel with
experiments there has been an intense theoretical effort in the
understanding of the mechanisms  underlying the
phenomenon\mycite{gro93}.  In fact a detailed and systematic study
is expected to provide fundamental properties of nuclear matter.
However, though many models are able to reproduce the data,  the
origin of multifragmentation has not been sufficiently clarified up
to  now\mycite{hir94}.

It seems nowadays established that multifragmentation  cannot be
explained by the same scenarios valid at lower energies.  The
characteristic time  ranges from  one hundred to few hundreds
$fm/c$ \mycite{dur94}.  Therefore it is a very fast process in
comparison with phenomena like fission or compound nucleus formation
whose  typical time is of the order of thousands $fm/c$.  Mechanical
instabilities\mycite{bau92} play a fundamental role  and they should
be taken into due account when the excited nuclear  system enters
the spinodal zone of nuclear matter Equation of State (EOS). The
latter is the region where the compressibility, $i.e.$  the
derivative of the pressure with respect to the density, becomes
negative and the system is unstable to small perturbations.

\par
The undoubted success of the statistical models\mycite{gro93,hir94},
in explaining some of the multifragmentation features, indicates
that the phase space dominates the population of the different final
channels. In such a fast process, the tendency of filling uniformly
the phase space cannot occur in each collisional  event. The system
has not enough time to explore, during the reaction,  the whole
phase space, as can  happen for compound nucleus collisions. It can
be expected, therefore, that the reaction dynamics is dominated by
the phase space only when the physical quantities are averaged over
large  sets of events. This assumption implies that the nuclear
dynamics in the  multifragmentation regime is irregular or chaotic
enough to produce, at least approximately, a uniform "a priori"
probability  to populate each region of the available phase space.
In particular, the formation of the final fragments must follow  an
irregular dynamics. This conjecture can be also inferred by the
large  event-by-event fluctuations observed experimentally on the
charge and  mass distributions\mycite{bau86}. \par Within this
scenario, however, it is still not clear at which stage  the
dynamics is dominated by a chaotic behaviour and  which is the
mechanism able to produce the expected phase space mixing. \par In
the present work the problem of fragment formation is approached in
a simplified way. It is assumed that the nuclear system after an
initial compression slowly expands entering into the spinodal
region. Some evidences of such a stage of the  reaction in central
heavy ion collisions have been found in  computer simulations based
on the BUU scheme\mycite{bau92,gro93}. The fragments are supposed to
emerge from this expanded nuclear system  through the unstable
density fluctuations of the spinodal zone. We study in details this
phase of the nuclear dynamics and show that it exhibits a chaotic
behavior. In section 1 we discuss a simple example to illustrate the
main features of deterministic chaos. The theoretical background for
the nuclear case is based on the Vlasov equation, which is
overviewed in section 2. In section 3 we discuss the numerical
method adopted to solve the Vlasov  equation on a lattice and study
the corresponding two-dimensional  Equation of State (EOS). The mean
field dynamics is investigated in section 4. The density profile of
the initially almost uniform nuclear system is followed until
fragments can be identified. The dynamics of this  process is
analyzed in order to investigate the possible non-linear and chaotic
behaviour, with methods analogous to the ones well known in the
theory of dynamical systems\mycite{lich83}. To this purpose, a
distance between "trajectories" is introduced, and the analysis in
terms of Lyapunov exponents is performed. In section 5 we discuss
the consequences of a non-linear and erratic dynamics for nuclear
multifragmentation. Finally in section 6 we draw our conclusions.
The results reported in this paper are a complete and detailed
review  of an investigation started along these lines in
refs.\mycite{bbr1,bbr2}.  \par

\section{Deterministic chaos in a simple example. }

In the last two decades the study of low-dimensional dynamical
systems has given a fundamental  contribution to the understanding
of the onset of irregular motion.  Due to the nonintegrability of
the sets of deterministic equations an erratic  and unpredictable
behaviour, {\it deterministic  chaos}, can follow. What was
previously considered as spurious noise has found  rigorous
mathematical and experimental foundations. This revolutionary
concept  has changed drastically our view of classical mechanics and
is revealing  important implications also in quantum physics.  For a
general review see ref.\mycite{lich83} The consequences of this {\it
new paradigm}  which is shading  light into apparently unrelated
different fields are difficult to overcast and will probably be an
important  guideline  of research for the next decades.  In this
work, we explore the dynamics of nuclear matter from this new
perspective. But before going to the discussion of the main topic,
we will consider the dynamics of a very simple system. This example
is used to illustrate the main features of  a chaotic unstable
system and the conditions under which  deterministic chaos can
occur.

Let us consider the two--dimensional problem described by the
hamiltonian
\be
\label {1}
{\cal H} \, =\, {1\over {2}}(p_1^2 + p_2^2)\, +\, q_1^2 + q_2^2
-2\mu q_1^2q_2^2
\ee
which corresponds to two coupled harmonic oscillators  with unit
mass. It can be easily seen that, for a given total energy $E$, the
available space has a fourfold structure, with four ``branches"
extending to infinity.  If $\mu\neq 0$, the dynamics of the particle
inside the potential is chaotic. In other words, it is impossible to
find an analytical solution for the equations of motion
corresponding to the  Hamiltonian (1). We have two degrees of
freedom and   only one constant of the motion, $i.e.$ the total
energy. This implies an extreme sensitivity  of the dynamics on the
initial conditions and a very irregular motion.

One can consider a particle starting its motion inside this
potential or  impinging on it from outside. In the latter case the
chaoticity of  the interaction region manifests itself in strong
irregularities,  at all scales, of the final observables - final
scattering angle, final internal excitation, etc. - as a function
of the initial conditions. We have the well known phenomenon of {\it
chaotic scattering}\mycite{ott93}.  An incoming particle can be
trapped for a long period inside the interaction region, where it
bounces erratically  to and fro  before escaping. The extreme
internal sensitivity  influences drastically the final result.

Classical chaotic scattering  is therefore intimately connected with
the presence of trajectories which remain trapped in the interaction
region for a time long enough that  chaos can set in.

These considerations can be readily extended to the case of
trajectories which starts directly from the interaction region and
which therefore will describe systems which are unstable and decay.
In particular,  the Hamiltonian (1) does not describe a proper
scattering situation, since the potential diverges at infinity.
However it can describe a system which escapes from the interaction
region after some trapping time. Since the system is chaotic, again
the trapping time and the direction of emission depend in a very
irregular way on the initial conditions. This point is illustrated
in fig.1.  For each trajectory starting from the  internal region,
one of the coordinates of the particle is recorded. The distance at
which we take this value is so large to be considered asymptotic.
The particle  cannot return anymore to the interaction region. In
the plot, the value of the final coordinate $q_f$ is reported as  a
function of the initial one $q_i$. It is important to notice  that
the considered set  of initial conditions lies on the same energy
surface.  The observed irregular structure is very similar to the
one usual encountered for the deflection function in chaotic
scattering systems. If chaoticity is strong enough, the distribution
of the trapping time is actually exponential\mycite{ott93}. More
formally, if we call ${\cal T}$ the average trapping time and $\tau
= 1/{\overline {\lambda}}$ the average rate of divergence  between
trajectories, being $\overline {\lambda}$ the average Lyapunov
exponent, one can write, according to T\'el\mycite{tel87}
\be
\label {2}
{\cal T}\, =\, \tau / (1 - {\overline D} )
\ee
where ${\overline D}$ is the average linear Hausdorff dimension of
the so-called {\it strange repellor}\mycite{ott93}. If ${\overline
D} > 0$, which is the condition for  a genuine chaotic scattering,
then ${\cal T} > \tau$.\par This condition can be used as a
criterium to argue if a system presents a chaotic dynamics  or not.
It is useful in applications, because it is usually numerically
difficult to assess the existence of positive ${\overline D}$ by an
extensive direct sampling of the  whole phase  space. Still it must
be used with caution, due to the unavoidable  uncertainties in the
numerical values of the Lyapunov exponents, and  therefore it can
give only an indication of a possible chaotic dynamics.\par

The analysis can be extended to models with many degrees of freedom.
In particular, for a fluid system the relevant degrees of freedom
can be identified with its linear eigenmodes, and in the harmonic
approximation a fluid is exactly equivalent to a set of harmonic
oscillators. Such a system, which corresponds to the case $\mu = 0$
in eq. (1), is integrable, since the energy of each mode is
conserved. Therefore, in this limit a fluid can undergo only a
regular dynamical evolution. If the amplitude of the perturbation is
large enough, the modes of the fluid can be coupled between each
other, as in eq. (1) for $\mu \neq 0$. In this case the problem is,
in general, non--integrable and chaotic dynamics  can be present.
\par If a uniform homogeneous fluid is initially in its spinodal
region, where the  compressibility is negative, it will
spontaneously "escape" towards a non--homogenous phase, where
droplets are formed.  However, before the final droplets are formed,
the fluid can spend enough time in the spinodal region to allow
non--linearity and chaotic motion to set in and dominate the time
evolution. The situation is quite analogous to the example of eq.
(1), where the particle escapes outside the interaction region but
has enough time to experience the chaotic dynamics present in the
inner sector of the potential, as illustrated in fig.1. If
chaoticity is really present, then the average Lyapunov exponent
$\overline{\lambda}$ must be positive and, as discussed above, the
criterium ${\cal T} > \tau$ should be satisfied. Here ${\cal T}$ is
the characteristic  time for droplet formation and $\tau =
1/\overline{\lambda}$ is the average time of divergence between
"trajectories" of the fluid. In the next sections we study - within
the semiclassical approximation - the dynamics of nuclear matter
inside the spinodal region. We will show,  following a reasoning
similar to the one discussed above, that nonlinearities and positive
Lyapunov exponents are found  inside the spinodal region. Therefore
deterministic chaos characterizes the formation of fragments.
\bigskip

\section{{\bf Theoretical background}}

\subsection {{\bf The Vlasov equation}}

In the multifragmentation energy range the dynamics is characterized
by an interplay between a purely mean field  evolution, typical of a
low-energy phenomenology (fusion, inelastic reactions), and two-body
collisions due to the partial relaxation of  the Pauli principle. A
possible way of incorporating both of these effects is the inclusion
of the residual interaction in models like TDHF, but this procedure
is difficult to carry on numerically in realistic
calculations\mycite{won}. For this purpose semiclassical methods
have been developed,  whose starting point is the Wigner transform
of the time-dependent Hartree-Fock (TDHF) equation for  the one-body
density matrix. By neglecting powers of $\hbar$ higher than
two\mycite{gre}, one obtains the usual Vlasov equation which reads

\be
\label {3}
{\partial f \over \partial t} + {{\bf p} \over m} {\bf \cdot \vec \nabla_r}
f - {\bf \vec \nabla_r} U[\rho({\bf r})] {\bf \cdot \vec \nabla_p} f = 0
\ee

\noindent
$f({\bf r},{\bf p}; t)$ is the Wigner phase-space distribution
function,  (${\bf r}, {\bf p}$) are the space and momentum
coordinates and $U$ is a self-consistent  single-particle potential
depending on the density  $~\rho=\int d{\bf p}~ f~$. When eq.(3)
includes a Nordheim-type collision integral on the right-hand side,
$I[f]$, it is generally  referred in literature as  $BUU$
(Boltzmann-Uehling-Uhlenbeck) or $VUU$
(Vlasov-Ueh\-ling-Uh\-len\-beck) or $LV$ (Landau-Vlasov) or VN
(Vlasov-Nordheim) equation. For a complete review, see
ref.\mycite{ber}. The collision integral reads

\be
\label{4}
I\lbrack f \rbrack ({\bf r}, {\bf p_1},t) =
\int d{\bf p}_2~ d{\bf p}_{1'}~ d{\bf p}_{2'}~(f_{1'} f_{2'} {\bar {f_1}}
{\bar {f_2}} - f_1 f_2 {\bar {f_{1'}}} {\bar {f_{2'}}}) \omega_{121'2'}
\ee
\noindent
where $f_j = f(s_j, t)$ is the phase-space occupation probability at
the location $s_j = ({\bf r}, {\bf p_j})$, and ${\bar {f_j}} = 1 -
f_j$  is the Pauli blocking factor of the  final states.
$\omega_{121'2'}$ is the microscopic transition rate for the
collision vertex $(p_1 p_2) \rightarrow (p_{1'} p_{2'})$, and is
related to the nucleon-nucleon scattering cross section. This
collision term describes the momentum changes of two interacting
particles during a collision with blocking factors forbidding
transitions leading to occupied final states.

We notice the formal identity of eq.(3) with the Liouville equation
of classical mechanics for a fluid of non-interacting particles.
This underlines its intrinsic classical character and guarantees
that quantum effects like the Pauli principle  are conserved. It can
be easily demonstrated that the Vlasov-Nordheim equation satisfies
the conservation laws of mass, momentum and energy. The presence of
an average potential $U$ dependent on the density introduces a
non-linear constraint on the solution of eq.(3). This is a common
point to all mean field theories.

Because of that, the Vlasov-Nordheim equation is difficult  to solve
numerically, in spite of its apparent simplicity. A large amount of
literature has been devoted to its numerical solution techniques in
the case of systems interacting through Coulomb fields like plasmas,
to which the Vlasov equation was  first applied.

Later on we will discuss the implications  of non-linearity on
physical processes like multifragmentation.

\bigskip

\subsection {{\bf Linear response for the Vlasov equation}}

The linearized version of the Vlasov equation has been successfully
applied to the study of small amplitude oscillations,  see
ref.\mycite{ccr,bur88}, and it represents the starting point for a
phase space approach to RPA. For small variations of the
distribution function around the  equilibrium solution $f_0 ({\bf
r}, p^2)$

\be
\label{5}
f({\bf r}, {\bf p},t) = f_0 ({\bf r}, p^2) + g({\bf r}, {\bf p},t)
\ee
\noindent
we can expand eq.(3) to the first order in g. A Thomas-Fermi
approximation for the ground state - which is obviously a solution
of eq.(3) - can be used. That is

\be
f_0 ({\bf r}, p^2) = {4 \over {(2\pi\hbar)^D}} \Theta(E_F - {p^2
\over {2m}})
\ee
\noindent
where D is the dimension of the physical space. Then one gets an
equation for the Fourier transform of $g({\bf r}, {\bf p},t)$ which
can be analytically solved and produces a dispersion relation for
the frequency $\omega$ for each wave number $k$. For stable systems
the linear response theory yields real values for the energies
$\omega=\omega (k)$  of the normal modes, which can be represented
by plane waves propagating in opposite directions. When the system
is unstable, some energies are complex and the imaginary part is
related to the time growth of instabilities. In both cases, the
dependence of  the energy $\omega$ on $k$ is linear in this
approximation, with a correction  factor depending on the equation
of state and the density of the  system, see
ref.\mycite{hei93,bur92,ccr}. We like to remind that the linear
response approximation has a limited range of validity, since it can
be applied only to small amplitude variations of the initial
reference state. Therefore its applicability to  processes like
multifragmentation, where large fluctuations are involved,  might be
not completely correct, as it will be discussed later.

\bigskip

\section{{\bf The model}}

\subsection {{\bf The lattice calculation}}

As previously explained in the introduction, we study the behavior
of a dilute nuclear system whose density fluctuations are growing
inside the spinodal region. In order to get numerically  robust
results, we solved the Vlasov-Nordheim equation on a lattice,  using
the same code of ref.\mycite{bur92}, but neglecting the stochastic
contribution to the collision integral. We divide the single
particle phase space into several small cells, each of volume $V^D =
\Delta r^D \cdot \Delta p^D$,  being $\Delta r$ and $\Delta p$ the
cell sidelengths respectively in  coordinate and momentum space.
Most numerical problems in calculating  the Vlasov evolution of
nuclear systems arise from the need to smooth the one-body density.
For this purpose we must employ a lattice with a big number of
small cells in order to get a nice paving of the phase space.
Typically $\Delta r$ is 1 fm or less, while $\Delta p$ should be
smaller than the Fermi momentum.  Of course the discretization
introduces some numerical error on the physical observables, but it
gets smaller and smaller as we decrease the cell size and increase
their number. Therefore the main limit of the  lattice method is the
memory resources of the computers and  the huge computing time
requested. For this reason, we performed all the calculations on a
two-dimensional lattice.

We have studied a fermion gas situated on a large torus  with
periodic boundary conditions, and its size is kept constant during
the evolution. The torus sidelengths are equal to $L_x = 51~fm$ and
$L_y = 15~fm$. We employed in momentum space 51x51 small cells of
size $\Delta p_x = \Delta p_y = 40~ MeV/c$, while in coordinate
space $\Delta x = 0.3333~fm$ and $\Delta y = 15~fm$, $i.e.$ we have
only one big cell on the $y$-direction.

The initial local momentum distribution was assumed to be the one of
a  Fermi gas at a fixed temperature $T$. We employ a local Skyrme
interaction which generates a mean field  $U[\rho] = t_0~(\rho /
\rho_0) + t_3~(\rho / \rho_0)^2$. The density $\rho$ is folded along
the $x$-direction with a gaussian of width $\mu = 0.61 fm$, in order
to give a finite range to the interaction. The parameters of the
force $t_0$ and $t_3$ have been chosen in order to reproduce
correctly the binding energy of nuclear matter at zero temperature,
and this gives $t_0 = ~-100.3~MeV$ and $t_3 = ~48~MeV$.  The
resulting EOS gives a saturation density in two dimensions equal to
$\rho_0 = ~0.55~fm^{-2}$ which corresponds to the usual
three-dimensional  Fermi momentum equal to $P_F =~260~MeV/c$.

Then a complete dynamical evolution is performed by subdividing the
total time in small time steps, each equal to $\Delta t = 0.5~fm/c$.

For more details concerning the mean field propagation on the
lattice and the exact calculation of the collision integral, the
reader is referred to ref.\mycite{bur92}.

\bigskip

\subsection {{\bf Two-dimensional nuclear matter equation of state}}

We calculate the nuclear matter equation of state (EOS) for an
homogeneous two dimensional system. We mainly follow the definitions
given in ref.\mycite{sau} for the three dimensional case.  For a gas
of particles interacting through a local Skyrme  single-particle
potential, $U[\rho]$, we calculate  the density of the free energy
{\bf {\it F}} and the corresponding pressure {\bf {\it P}}. They
read

\be
\label{6}
F = H - TS
\ee

\be
\label{7}
P = - F + \rho ~(\mu + U[\rho])
\ee

\vskip 20pt
\noindent
In eq.(7) {\bf {\it H}} is the spatial energy density, and it is
given by summing the kinetic energy density {\bf {\it K}} and the
potential  energy density ${\bf {\it E = \int U[\rho] d\rho }}$.
$\mu$ denotes the chemical potential, T is the temperature and S is
the density of entropy, for which we use the standard definition for
noninteracting particles\mycite{sau}.

If we define the parameter

\be
\eta = {\mu \over T}
\ee

\vskip 20pt
\noindent
the single-particle density, the kinetic energy density and the
entropy density can be expressed through the Fermi integrals $J_0$
and $J_1$

\be
\label{8}
\rho = {{2 m} \over {\pi \hbar^2}}~ J_0(\eta)
\ee

\be
\label{9}
K = {{2 m} \over {\pi \hbar^2}}~ J_1(\eta)
\ee

\be
\label{10}
S = {{4 m} \over {\pi \hbar^2 T}}~ J_1(\eta) - \eta \rho
\ee

\vskip 20pt
\noindent
where the Fermi integral is given by

\be
\label{11}
J_\nu(\eta) = \int {{\epsilon^\nu d\epsilon}
\over { 1 + e^{{\epsilon \over T} - \eta }}}
\ee

\vskip 20pt
\noindent
Since the integral $J_0$ is analytical, the chemical potential $\mu$
is readily calculated from the single-particle density $\rho$

\be
\label{12}
\mu = T ~ log \lbrack exp( {{\pi \hbar^2 \rho} \over{ 2mT}})
- 1 \rbrack
\ee

\vskip 20pt
\noindent
{}From the above definitions, the expressions for the density of the
free  energy and the pressure are

\be
\label{13}
F = - {{2m} \over {\pi \hbar^2}}~ J_1(\eta) + E + \mu \rho
\ee

\be
\label{14}
P = {{2m} \over {\pi \hbar^2}}~ J_1(\eta) - E + \rho ~U[\rho]
\ee
In Fig.2 we show the two-dimensional nuclear matter equation of
state (EOS), calculated for the Skyrme interaction we discussed in
the previous subsection. In the part a) of the figure, the free
energy per particle is  displayed along the isotherms for different
values of the temperature $T$ as function of the density, while in
part b) the corresponding thermostatic  pressure is shown. This
recalls the qualitative features of a  classical Van der Waals gas.
We notice that the general trend of a two-dimensional EOS does not
differ appreciably from the realistic three-dimensional
case\mycite{sau}.

It has to be noticed that the pressure $P$ can be also calculated
by the thermodynamical relationship  $P=-\rho^2 {\partial F\over
\partial \rho}$, being  $F$ the free energy per particle. It can be
readily verified that this procedure  gives the same expression of
eq.(16). The latter result indicates that in  our particular simple
model of nuclear matter, the thermodynamical  relationships are
exactly satisfied. This is a consequence of the simple  form adopted
for the interaction.

The spinodal zone can be identified in fig.2b)  as the region where
the compressibility , {\sl i.e.} the slope of the pressure $P$  as a
function of the density $\rho$, is negative. At increasing
temperature the density interval where ${\partial P \over \partial
\rho}$ is negative becomes smaller and smaller. Finally at the
critical temperature $T_c$ for the liquid-gas phase transition the
spinodal zone reduces to a point where the corresponding isotherm
has an inflection. For $T>T_c$  the pressure $P$ is a monotonic
increasing function of $\rho$. From the figure one can  deduce  $T_c
\sim 16~ MeV$.

\section{ Mean field dynamics }

In this section we want to study in detail the response of nuclear
matter to small perturbations. We neglect for the moment the
collision term I(f)  and solve numerically only the Vlasov equation.
If non-linear terms are  negligible, linear response theory gives an
accurate description of the time evolution. However we will see that
this is not always the case.

Let us consider an initial  sinusoidal perturbation along the
$x-$direction in the  density profile. That is
\be
      \rho(x) = [~1 + \delta sin(k x) ]~ \overline{\rho} ~,
\ee
with
\be
       k = {2\pi n_k\over {L_x}}   ~ ,
\ee
\noindent
being $L_x$ the size of the torus along the $x$-direction and
${\overline\rho}=\int dx \rho(x) / L_x $. We study the time
evolution of this profile considering a small amplitude for the
perturbation, $i.e.$  $\delta~= ~0.01$. In fig.3 and 4 we show this
evolution for  ${\overline \rho}/\rho_0 = 0.8$ for the cases
$n_k=4,5$.  In these figures  the mode initially excited is damped
in  a time  range which is around 40 $fm/c$.  This is the usual
Landau damping of Fermi liquid theory which is a well known linear
phenomenon. Actually, since we are working in a discrete lattice, we
have also a small initial excitation of a subharmonic of higher
order, $i.e.$   $n_k=12$ and $n_k=25$, which is also damped very
rapidly. In a previous paper\mycite{bbr1}  we had shown a similar
damping but at normal density. In that case the damping is slower
and occurs within  70 $fm/c$. The Landau damping  for nuclear matter
in two dimensions is an important issue by itself  and will be
discussed in detail in a forthcoming paper\mycite{landau}.

The behavior shown in the two previous figures indicates that
initial perturbations evolve linearly in time outside the  spinodal
zone. Moreover a small difference in the initial conditions does not
produce a different dynamical evolution, as we will discuss later.

The situation is  completely different if one analyses the time
evolution  inside the spinodal zone, $i.e.$ for
${\overline\rho}/\rho_0 < 2/3 $. In fig.5 we display the evolution
of three initial perturbations at ${\overline {\rho}}/\rho_0 = 0.4$.
The small initial modulations evolve and form fragments with
different shapes and sizes.  In our case, due to the fact that we
are considering a two-dimensional torus of nuclear matter, no real
fragmentation occurs and we call "fragments"  the macroscopic
structures of the density profile.  The behavior shown in fig.5 is
typical of a dynamical system in a chaotic  region, where very small
initial  perturbations are rapidly amplified and distorted.   The
relevant characteristic   is that {\it the density profiles do not
increase rapidly in their amplitude only, they also change their
shape strongly during the time evolution}. These features are strong
indications of a  non-linear regime.

In fig.6  we show the  power spectra corresponding to the modes
involved in fig.5. These spectra are different in each case and
illustrate  a population of more and more modes as the time
evolution goes on.   Hence the behaviour  of the system for
${\overline\rho}/\rho_0=0.4$ show a sensitive dependence on the
initial  conditions and an unpredictability of the evolution for a
given initial  perturbation. Small initial differences will produce
large final deviations\mycite{nota}.  A similar feature was
illustrated in ref.\mycite{bbr1} for ${\overline\rho} /\rho_0=0.5$.
As we will clarify in the following this behavior is typical inside
the  spinodal region and does not depend on the particular choice of
varying the initial conditions. In the present case we have changed
the initial $n_k$,  but we could have modified either the average
density or the amplitude  of the initial perturbation.

The sensitivity to the initial conditions occurs, in the dynamical
evolution, together with a dominant role of the non-linear terms in
the Vlasov equation. This is proved by the time evolution of the
strength of the most important modes as illustrated in fig.7. In the
latter the numerical calculations are reported as open squares in
steps of 5 $fm/c$. Linear fits of the initial evolution are shown
for comparison.  It is evident that the growth is linear only in
the first stages of the evolution while it strongly deviates from
the  fits after $\sim 25 fm/c$. In general one cannot trust linear
response theory after the very first dynamical stages. It is true
that   some modes follow a linear evolution for a longer time,
however in general they do correspond to very small wavelengths.
Therefore they do not affect the size of the main emerging fragments
but only their surface modulations. One can consider $\lambda$/2 as
the  diameter of the nuclear fragment formed. In concluding this
section one can say that when the system lies inside the spinodal
region many modes are coupled and they interact strongly among each
other. The final fragments are the result of this dynamics which
goes beyond the linearized equations discussed in section 2.2.

\bigskip \bigskip \bigskip

\subsection{ \bf{Calculation of the largest Lyapunov exponent }}

The arguments discussed so far provide a strong indication that the
dynamics is chaotic beyond the spinodal line, but they do not
provide a real proof that this is so. Therefore in this section  we
present numerical calculations which define the degree of
chaoticity according to the usual techniques adopted in the study of
low-dimensional dynamical systems\mycite{lich83}. The discussion
follows the same lines of ref.\mycite{bbr1,bbr2} but gives a more
complete and detailed overview.

A way to characterize quantitatively the dynamics in a chaotic
regime is  by means of the rate of divergence of two nearby
trajectories. The mean rate is usually called largest Lyapunov
exponent\mycite{lich83} and it is calculated by means of the
expression
\be
{\overline{\lambda}} \, = \, \lim_{t\to\infty}\, \lim_{d_0\to0}\, \lambda(t)
\ee
where
\be
\lambda(t)\, =\, {\log (d(t)/d_0) \over t}  ~.
\ee
In the latter $d(t)$  is the distance  between two phase space
trajectories, along an unstable direction. In our notation  $d_0 =
d(0)$. We have chosen as metric the following one
\be
d(t) \, =\, \sum_i \, | \rho^{(1)}(x_i, t) - \rho^{(2)}(x_i, t)
|/N_c ~,
\ee
\noindent
where the index $i$ runs over the $N_c$ cells in ordinary space, and
$\rho^{(1)}(x_i)$, $\rho^{(2)}(x_i)$ are the densities in the cell
$x_i$  for the trajectories $1$ and $2$ respectively. The definition
of eq. (21) represents  the difference between the two density
profiles and is sufficient  for our purpose, although possible
differences in momentum space  are averaged out. It should include
the contribution of all the unstable modes which dynamically grow up
during the evolution.

In fig.8 we show the time evolution of $\lambda(t)$  for various
average densities. Inside the spinodal region $\lambda(t)$ converges
to a limiting value $\overline{\lambda}$,   represented by a dashed
line, which is just the largest Lyapunov  exponent. This convergence
occurs only for a limited time interval, which is the one needed for
fragment formation. As it will be discussed below this  is because
we are describing a transient phenomenon. A different behavior is
observed when the system evolves outside the  spinodal region. In
fact in this case the trajectories do not diverge exponentially and
$\lambda(t)$ goes  slowly towards zero as time increases. This
result is analogous to the general case occurring  in simple
dynamical  systems, when $\overline{\lambda}$ finite and positive
distinguishes a  chaotic behavior from a regular one where
$\overline{\lambda}=0$. The values of  $\overline{\lambda}$   at a
temperature T=3 MeV are  reported on the second column of table 1.
Some of the  features here discussed were already presented in
ref.\mycite{bbr1} where actually the evolution was followed for a
shorter time.  For the calculations presented in this paper we
improved the algorithm used and we could follow the time evolution
for a longer period maintaining an error in the energy conservation
smaller than 1$\%$. Now one can see that, inside the spinodal zone,
$\lambda(t)$  remains around the plateau - which indicates the value
of the largest  Lyapunov exponent - only for a limited time range.
This fact has  a simple explanation. In fact, the time at which the
plateau stops corresponds to the primary fragment formation. Hence
after  fragments are formed, the different modes interact  weaker
than before  among each other, because along the $x$-direction there
are  many points where the density is almost zero. These fragments
stop increasing their size and start to move along the
$x$-direction. They eventually collide against each other and
therefore,  in effect, the fragment configuration continues to
change. However this  dynamical regime is different from the
previous one and we will not consider  the time evolution following
the plateau. It is important to notice that estimates of the
expansion time in the BUU framework\mycite{gro93} are surely larger
than 100 $fm/c$. Hence the expansion process is expected to hardly
affect the dynamics of fragment formation, while it could have some
effects on the successive stages of the reaction. A realistic
investigation of this following regime can be approached only by a
fully dynamical simulation and it is beyond the scope of the present
work.

Now we discuss the reliability of the numerical estimates for
$\overline{\lambda}$ in our calculations. In fig.9 we show the
dependence of $\overline{\lambda}$ on the initial distance $d_0$ for
average densities $\overline{\rho}/\rho_0$=0.4, 0.5.   There is a
strong dependence of  $\overline{\lambda}$ on $d_0$ for large
values, which however saturates for values smaller than  $\sim 5.
\cdot 10^{-3} fm^{-2}$ In the calculations displayed in fig.8  we
actually used a value  $d_0 \leq  10^{-4} fm^{-2}$ ,as in
ref.\mycite{bbr1}, which lies  in the  saturation zone and therefore
assures a stable numerical result.  Last but not least, we also
checked that $\overline{\lambda}$'s do not change varying the size
of the  torus or changing the time step of the
integration\mycite{bbr2}.  In fig.10 the dependence of
$\overline{\lambda}$ on the initial $n_k$ is displayed for
$\overline{\rho}/\rho_0$=0.4,0.5.  Even in this case the
insensitivity to the variation of the wave number assures that the
value of the Lyapunov  exponent  is numerically stable. It is
important to notice that $\overline {\lambda}$ is different from the
growth rate calculated in the linear response theory, where one
should expect a linear dependence with the largest mode having the
fastest growth. Only if the system is linearly unstable the Lyapunov
exponent and the growth rate do coincide. This is not true in our
case where the latter is smaller by almost a factor of two.

Up to now the momentum distribution adopted was always that one of a
Fermi gas  at temperature T=3 MeV. We varied also this temperature
to see the effect  on $\overline{\lambda}$. In fig.11 the dependence
of $\overline{\lambda}$ on the temperature is shown. We found a
small dependence on the temperature which indicates that  thermal
motion seems to reduce only slightly the degree of chaoticity. This
is due to the shrinking of the spinodal zone at increasing
temperatures, where vaporization processes start to compete with
instabilities.

The calculations discussed up to now refer only to the Vlasov
equation. One could wonder how much the inclusion of  two-body
collisions can modify the discussed scenario. To this purpose, we
solved the Vlasov-Nordheim equation numerically and calculated the
Lyapunov exponents also in this case. In fig.12 we show the
comparison of  ${\lambda(t)}$ calculated with and  without two-body
collisions. The figure shows  that the inclusion of the collision
integral (4) in the r.h.s  of eq. (3) influences only  slightly
$\overline{\lambda}$ reducing its value.  This reduction is of the
order of 30$\%$ for $\overline{\rho}/\rho_0$=0.5, but only of 10$\%$
for $\overline{\rho}/\rho_0$=0.4\mycite{bbr1}.  It is not difficult
to understand this result by considering  the diluteness of the
fermion gas inside the spinodal region.

\newpage

\subsection{ \bf{ Time scales}}

In this section we discuss the interplay of the  different time
scales  occurring in fragment formation.

In fig.13 we show the comparison of the Lyapunov time  $\tau_{chaos}
=1/\overline{\lambda}$ and the time needed to form  the primary
fragments $\tau_{frag}$, corresponding to the time end of the
plateau. The former is always smaller than the latter. It is
important to notice that $\tau_{frag}$ is analogous to the trapping
time $\cal T$ discussed in section 1. Hence according to the
criterium there introduced for the simple example, the relation
$\tau_{frag} > \tau_{chaos}$ implies that, during the interval of
time in which fragments are formed, deterministic chaos can fully
develop and characterizes the clusterization.

An additional evidence of that can be also seen in fig.7, where
some modes, despite the overall growth of density fluctuation, can
have amplitudes oscillating in time. This behaviour is strongly
reminiscent of the erratic bouncing of the particles inside the
potential of the example discussed in section 1.  Of course the
dynamics is followed in a finite time interval and therefore in our
case  chaoticity is only a transient phenomenon. Poincar\'e maps
like in ref.\mycite{moo93} cannot be drawn. We believe however that
the evidence found are strong enough to justify at least a  strong
analogy of the above discussed dynamics with the more  familiar
low-dimensional chaoticity in closed and scattering systems
\mycite{lich83}.

It can be instructive to compare the characteristic times
$\tau_{chaos}=\hbar/{\overline \lambda}$, which defines the time
scale of the divergence between mean field trajectories, with the
single particle characteristic time $\tau_{sp}=\hbar/E_F(\rho)$,
being $E_F$ the Fermi energy at the given density. This is done in
table 1 for a set of densities and for a temperature T=3 MeV. One
can see that for densities $\rho \leq 0.4 \rho_0$ the divergence
time is smaller than the single  particle time. In other words the
motion associated with the mean field is faster than that of the
particles moving inside. Therefore in this region the  notion itself
of mean field ceases to have validity, being the gas too dilute.

\bigskip \bigskip \bigskip

\section{ Chaoticity implications for multifragmentation }

In this section we discuss what are the consequences of a chaotic
dynamics in the spinodal zone  for the final fragments formation. In
our schematic description we are  neglecting the initial compression
induced by the heavy ion collision which likely leads the composite
system well inside the spinodal zone.  However one could simulate
the uncertainty in the initial conditions by considering in  our
model  a small and random initial perturbation in the density
profile. This is also another way of studying the response of our
system inside the spinodal region. In particular we start the time
evolution giving to each cell along the $x$-direction a random shift
around the average density whose strength is $1\%$ of the average.
Also with this initialization the time evolution is non-linear and
show the same features already discussed in the previous section for
a  sinusoidal shape. This behaviour is shown in figs.14 and 15 where
the  time evolution of two random initialized events are compared by
displaying the density  profiles and the power spectra at different
times. In fig.14 the scale at t=0 is different to magnify the random
initial shape. The power spectra of this shape is rather flat as
fig.15 illustrates. During the first evolution a wide range of modes
are privileged due to the finite interaction range and then among
these an erratic and hardly predictable mixing occurs. The dynamics
of the single modes  exhibits the same non-linear trend illustrated
in fig.7 \mycite{tao94}. We have checked that calculating the
Lyapunov exponent with a random initialization one gets values
identical to those shown previously.

The initial random shape is on one hand more realistic, simulating
the  missing dynamics, and on the other hand it gives the
opportunity to  sample many initial conditions. In this perspective
we performed  108 runs for an average density
$\overline{\rho}/\rho_0$=0.5 and extracted the fragment size
distribution out of each one.  In our case we do not have real
fragments but a natural  criterium is to consider those contiguous
cells whose density exceeds a  threshold value as forming a cluster.
We adopted a freeze-out time which is slightly larger than that
shown in fig.13 and defined as $\tau_{frag}$ and is equal  to 120
$fm/c$. This was necessary because the dynamics is somehow slower
when many modes are coupled since the beginning. Therefore the time
for fragment formation shown in fig.13 should be considered  a
minimum estimate. In fig.16 we show  the average fragment size
distribution $P(s)$ for  three different threshold cuts equal to 5,7
and 10 $\%$ of the normal density $\rho_0$.  As one can see the
three distributions  are rather large and do not differ
significantly between each other.  In the same figure two fits for
each distribution are shown. The dashed curve is an exponential fit,
$i.e.$ $P(s) \sim e^{-\tau s}$, while the full curve corresponds to
a power law fit, $i.e.$  $P(s) \sim s^{-\tau}$. The corresponding
$\tau$ and the relative errors are reported in the figure.  The fits
shown in the figure were performed  up to a maximum value
$s_{max}=20$.  In table 2 we report the values  obtained, together
with the  respective $\chi^2$,  by  fitting the three distributions
up to $s_{max}=20,30$. In all cases the power law fits have  a
smaller  $\chi^2$ than the exponential ones.  Though the difference
is  small, a power law fit always seems more appropriate. The value
of the $\tau_{power}$ which one can extract from the power law fits
oscillates around 2.  In a preliminary analysis of this kind we took
into  account 100 events considering a freeze-out time smaller and a
larger cut\mycite{bbr2}. However a  power law fit gave also in that
case a similar value for $\tau_{power}$. Thus the more refined
present analysis does not change the previous results.

One could be tempted to compare these calculations to those
published recently concerning experimental data and theoretical
considerations \mycite{gro93,ala94}. In fact, the large
distributions shown in fig.16  remind those observed  experimentally
and the exponents found are also very close  to those of percolation
at criticality\mycite{bau86}. Actually, these similarities could be
accidental. Our model is at the  moment too schematic and the
criteria adopted are rather arbitrary  to draw final conclusions.
However, we think it is very important  the fact that a
deterministic chaotic behavior allows for final large  distributions
which resemble those found in experimental data.  On the contrary we
note that within a linear unstable  evolution\mycite{ccr,hei93}  the
most unstable mode has the largest growth rate and wins always over
the others. Therefore,  in the linear regime, the path followed by
the system in its  dynamical evolution from a random small
initialization towards a macroscopic fragmentation is always the
same, apart from a very small spreading width. Hence one obtains a
strongly peaked and gaussian size distribution. Such a kind of
multifragmentation has not yet been observed experimentally.

In concluding this section we claim that a deterministic chaotic
mechanism  is the most natural explanation for the
multifragmentation data. In a chaotic regime even selecting events
with the same impact parameter, energy, temperature,  etc. the small
uncertainties which will always be present would lead  inevitably to
very large fluctuations with mass and charge distributions ranging
from an exponential to a power law.  Chaoticity does not exclude
phase transitions or statistical hypotheses. On the other hand it
seems likely to be a general feature underlying  these scenarios.

\bigskip \bigskip \bigskip

\section{ Conclusions}

The growth of density fluctuations, in the spinodal region of
nuclear matter EOS, has been advocated as the mechanism of fragment
formation in the multifragmentation process observed in heavy ion
collisions. According to this scenario, the nuclear fragment
formation is similar to the process of droplet formation in an
over-saturated classical vapour. We have studied the dynamics of
such a  process by solving the Vlasov--Nordheim equation numerically
on a lattice. The nuclear system in the spinodal region was
schematized by a  two--dimensional fluid confined inside a fixed
torus. The initially  homogeneous system was perturbed by adding a
small density fluctuation and the dynamics was followed up to the
time of fragment formation, when several well separated density
humps are apparent. Several  characteristics, typical of chaotic
dynamics, have been identified and  studied. \par\noindent {\it i})
Strong sensitivity to the initial conditions is apparent in the
evolution of the system. Small variations of the initial conditions
lead to  large deviations in the final pattern of the density
profile and  therefore of the fragment distribution.\par\noindent
{\it ii}) The amplitudes of the modes increase, in general,
non--exponentially, with possible oscillating behaviour, at least
for the wavelengths relevant for the fragment formation ( $\lambda >
6 fm$ ). This indicates that the modes are coupled and the dynamics
is highly non--linear.\par\noindent {\it iii}) The analysis in terms
of Lyapunov exponents gives values different from zero and
independent from the wavelength of the initial perturbation. This is
in sharp contrast with the expectations for a  linear dynamics, for
which the inverse Lyapunov exponent coincides with the growth rate
of the fluctuation and is proportional to the momentum of the
mode.\par\noindent {\it iv}) The growth rate of the density
fluctuations is longer than the  inverse average Lyapunov exponent,
and therefore chaotic dynamics has time  to develop during the
process of fragment formation. This criteria of identifying chaotic
dynamics has been illustrated with a simple example in Sec. 1 \par
Points {\it i--iv} give strong evidences of a chaotic behaviour of
the  dynamics in the fragment formation process. Of course such a
process takes place in a finite time interval, of the order of 100
$fm/c$, and therefore the traditional method of drawing Poincar\'e
maps to identify possible chaotic dynamics, cannot be used here. The
situation, however, is similar to the case of chaotic scattering,
and we believe  that the evidences are strong enough to justify the
conclusion of a chaotic dynamics of the process.\par This conclusion
gives support to the statistical models used to analyze the
multifragmentation data, since chaotic dynamics entails a  tendency
of the system to fill uniformly the available phase space. In fact,
if the dynamics is chaotic, strong fluctuations are expected from
one  collision event to another, each one ending in a different
region of the available phase space. On the average, therefore, if
chaoticity is strong enough, the population of the final channels
will be dominated by phase space.\par Finally, a preliminary
analysis of the fragment size distribution shows an approximate
power law. It has to be pointed out that the presence of chaotic
dynamics does not exclude the possibility that the system passes
through a phase transition. On the other hand a power law does not
necessarily imply a phase transition.

More detailed and realistic investigations have to be performed in
order to confirm this appealing scenario. It should be also clear
that we have {\it not presented a new model}, but  rather  {\it a
new perspective } in nuclear dynamics which seems more appropriate
and promising.

\bigskip
\noindent
We would like to thank M. Ploszajczak, D.H.E. Gross, Ph. Chomaz,
J. Randrup, X. Campi and M. Di Toro for stimulating discussions.

\bigskip
This work was partially supported by the Human Capital and Mobility
Program of the European Community contract no. CHRX-CT92-0075.

\newpage

\begin {table}

\begin{center}
\begin{tabular}{|cccc|}
\hline
    ~                 &     ~          & ~        &        ~  \\
{}~  ${\overline \rho}/\rho_o$ ~ &
 ~$\overline{\lambda}$  ~    & ~$\tau_{chaos}$ ~&  ~$\tau_{sp}$ ~  \\
    ~~~~~~                 &~
$c/fm$    ~          &~$fm/c$       ~&  ~$fm/c$      ~  \\
    ~                 &     ~          & ~        &        ~  \\
\hline
    ~                 &     ~          & ~        &        ~  \\
    ~ 0.7~            & ~ 0.
     ~ & ~$\infty$~       &  ~ 7.8 ~                \\
    ~ 0.6~            & ~ $2.8 \cdot 10^{-2}$
  ~ & ~28.6 ~          &  ~ 9.1 ~                \\
    ~ 0.5~            & ~ $6.8 \cdot 10^{-2}$
  ~ & ~14.7 ~          &  ~ 10.9 ~               \\
    ~ 0.4~            & ~ $9.6 \cdot 10^{-2}$
   ~ & ~10.4 ~          &  ~ 13.6 ~               \\
    ~ 0.3~            & ~ $0.10$
 ~ & ~10. ~           &  ~ 18.1 ~               \\
    ~ 0.2~            & ~ $0.10$
   ~ & ~10.  ~          &  ~ 27.2 ~               \\
    ~                 &     ~          & ~        &        ~  \\
\hline
\end{tabular}
\caption{For a temperature T = 3 MeV, the Lyapunov exponent
$\overline{\lambda}$, as calculated in the text, and the
corresponding characteristic time $\tau_{chaos}= 1
/\overline{\lambda}$ are reported. For comparison the single
particle time $\tau_{sp}=\hbar/E_{F}(\rho)$ for various values of
the average  density outside and inside the spinodal region is
shown.}
\label {tab1}
\end{center}
\end{table}

\begin {table}

\begin{center}
\begin{tabular}{|c|cc|cc|}
\hline
    ~$s_{max}=20$~ &     ~
    &                    & ~                   &            ~  \\
    ~              &     ~
     &                    & ~                   &            ~  \\
    ~ ${({\overline  \rho}/\rho_0)_{cut}}$
   ~    & ~$\tau_{power}$ ~
& ~$\chi^2_{power}$~ & ~$\tau_{expon} $ ~  & ~$\chi^2_{expon}$ ~  \\
 ~  ~&     ~
&                    & ~                   &            ~  \\
\hline
    ~             &     ~
     &                    & ~                   &            ~  \\
    ~  0.05 ~     &  ~ 1.48$\pm$0.16~
  &  ~ 0.087~          & ~   0.10$\pm$0.01 ~  &   0.12     ~  \\
    ~  0.07 ~     &  ~ 2.17$\pm$0.49~
  &  ~ 0.80~           & ~   0.15$\pm$0.04 ~  &   0.96     ~  \\
    ~  0.10 ~     &  ~ 2.17$\pm$0.34~
&  ~ 0.61~           & ~   0.15$\pm$0.03 ~  &   0.86     ~  \\
    ~             &     ~
   &                    & ~                   &            ~  \\
\hline
\end{tabular}

\begin{tabular}{|c|cc|cc|}
\hline
     ~$s_{max}=30$~ &     ~
&                    & ~                   &            ~  \\
     ~              &     ~
    &                    & ~                   &            ~  \\
    ~  ${({\overline  \rho}/\rho_0)_{cut}}$ ~
  & ~$\tau_{power}$ ~   & ~$\chi^2_{power}$~
 & ~$\tau_{expon} $ ~  & ~$\chi^2_{expon}$ ~  \\
  ~ ~ &     ~               &
        & ~                   &            ~  \\
\hline
    ~             &     ~
   &                    & ~                   &            ~  \\
    ~  0.05 ~     &  ~ 0.96$\pm$0.13~
  &  ~ 0.62~           & ~   0.05$\pm$0.01 ~  &   0.73     ~  \\
    ~  0.07 ~     &  ~ 1.62$\pm$0.22~
  &  ~ 1.770~          & ~   0.09$\pm$0.01 ~  &   2.13     ~  \\
    ~  0.10 ~     &  ~ 2.10$\pm$0.15~
   &  ~ 1.010~          & ~   0.12$\pm$0.01 ~  &   1.32     ~  \\
    ~             &     ~
   &                    & ~                   &            ~  \\
\hline
\end{tabular}
\caption{ The values of the exponential and power law fits to the
fragment distribution shown in fig.16 are shown. The fits were
performed up to a maximum $s$ value $s_{max}= 20,~30$.}
\label {tab2}
\end{center}
\end{table}

\newpage
\begin{figure}
\label{f1}
\noindent
\caption{ Final value $q_f$ of the coordinate $q_1$ for a given set
of initial  conditions, defined by the total energy E = 3.9 , and
the constraints   ${1\over 2}p_1^2 + q_1^2 = 3.7$, $q_2 = 0$. The
initial value $q_i$ of the  coordinate $q_1$ is taken randomly in
the corresponding allowed interval. }
\end{figure}

\begin{figure}
\label{f2}
\noindent
\caption{
The free energy per particle $F$ and the pressure $P$ are
respectively shown in part (a) and (b). Both quantities are
calculated as function of the density $\rho/\rho_0$ and for
different temperatures $T$.} \end{figure}

\begin{figure}
\label{f3}
\noindent
\caption{Time evolution of the density profile (a) for an average
density ${\overline \rho}/\rho_0=0.8$. The initial harmonic
perturbation is characterized by a node number $n_k = 4$. The
corresponding power  spectrum is shown in (b). } \end{figure}

\begin{figure}
\label{f4}
\noindent
\caption{The same as fig.3 but for $n_k = 5$.}
\end{figure}

\begin{figure}
\label{f5}
\noindent
\caption{Time evolution of the density profile for an average
density ${\overline \rho}/\rho_0=0.4$.  The examples displayed above
correspond to three different initial conditions, obtained by
changing the node number $n_k = 3,\ 4,\ 5$.} \end{figure}

\begin{figure}
\label{f6}
\noindent
\caption{Power spectra corresponding to the density profiles shown
in fig.5. }
\end{figure}

\begin{figure}
\label{f7}
\noindent
\caption{Time evolution of the strength of the main modes excited in
the profile of fig.5 for $n_k = 5$ is shown as open diamonds in
steps of 5 $fm/c$. The linear fits over the first four points are
drawn as full lines.  The corresponding wavelengths are also
reported.}
\end{figure}

\begin{figure}
\label{f8}
\noindent
\caption{ The quantity $\lambda(t)$ is shown as a function of time
for different values of the average density reported also in the
figure. The dashed line correspond to the largest Lyapunov exponent
$\overline\lambda$. The full curve is only to guide the eye.}
\end{figure}

\begin{figure}
\label{f9}
\noindent
\caption{The value of $\overline\lambda$ is shown as a function of
the initial  distance $d_0$ for different values of the initial
density  ${\overline\rho}/\rho_0=0.4,0.5$. See text for further
details.} \end{figure}

\begin{figure}
\label{f10}
\noindent
\caption{The value of $\overline\lambda$ is shown as a function of
the node number $n_k$ characterizing the initial harmonic
oscillation. The squares  correspond to the initial average density
${\overline\rho}/\rho_0=0.4$, while the diamonds correspond to
${\overline\rho}/\rho_0= 0.5$. The dashed lines represent the
average value.} \end{figure}

\begin{figure}
\label{f11}
\noindent
\caption{The value of $\overline\lambda$ is shown as a function of
the  temperature T for an initial average density
${\overline\rho}/\rho_0=0.5$.} \end{figure}

\begin{figure}
\label{f12}
\noindent
\caption{Time evolution of $\lambda(t)$ for an average
initial density ${\overline\rho}/\rho_0=0.5$ respectively without
(crosses) and with the collision integral I[f] (squares).}
\end{figure}

\begin{figure}
\label{f13}
\noindent
\caption{The Lyapunov time $\tau_{chaos}=1/{\overline\lambda}$
(diamonds)  is shown in comparison to the time requested by the
system in order to form fragments  $\tau_{frag}$ (squares) at the
various average densities. The initial condition is an harmonic
perturbation of the initial density profile.} \end{figure}

\begin{figure}
\label{f14}
\noindent
\caption{Time evolution of two random initial density profiles. The
average density is ${\overline \rho}/\rho_0=0.5$. Please note the
magnification of the initial scale.} \end{figure}

\begin{figure}
\label{f15}
\noindent
\caption{Power spectra corresponding to the density  profiles
displayed in fig.14. Also in this case the initial scale is
magnified.}
\end{figure}

\begin{figure}
\label{f16}
\noindent
\caption{Fragment size distributions $P(s)$ obtained by considering
108 events at an average density ${\overline \rho}/\rho_0=0.5$.
$P(s)$ is normalized to the total number of events considered. Each
event is generated through a random initialization of the kind shown
in fig.14. The three panels correspond to the different density cuts
chosen in order to define the fragments. In the figure are also
drawn the power law fit (solid line) and the  exponential one
(dashed line). The maximum value for the fit was $s_{max} = 20$. The
values of the fitted parameters  with the relative errors are
reported too. See text for further details. } \end{figure}


\bigskip

\begin{thebibliography}{99}

\bibitem{hir94}
For a complete review see :


Proceedings of the international conference "
Gross properties of nuclei and nuclear excitations",
Hirschegg, Austria, eds. H. Feldmeier and W. N\"orenberg,
GSI Darmstadt 1994 and references therein.


Proceedings of the {\it XXXII Winter Meeting on Nuclear Physics},
January 24-29, 1994, Bormio, Italy, edited by I. Iori,
Ricerca Scientifica ed Educazione Permanente.

Proceedings of the {\it 10th Winter meeting
on Nuclear Dynamics}, January 15-22, 1994, Snowbird, Utah, USA,
eds. W. Bauer and J. Harris, World Scientific, in press.

\bibitem{gro93}
For a general discussion see:

D.H.E. Gross, {\it Nucl.\ Phys.}
{\bf A553},  175c (1993) and references therein ;

\bibitem{dur94}

M. Aboufirassi {\it et al.}, Proceedings of the {\it XXXII Winter
Meeting  on Nuclear Physics}, pag.1, January 24-29, 1994, Bormio,
Italy



\bibitem{bau92}
L. Moretto, K. Tso, N. Colonna, G. Wozniak , {\it Phys.\ Rev.\
Lett.} {\bf 69}, 1884 (1992);  W. Bauer, G. Bertsch and H. Schultz,
{\it Phys.\ Rev.\ Lett.} {\bf 69}, 1888 (1992).


\bibitem{bau86}
W. Bauer, {\it Phys.\ Rev.} {\bf C38}, 1297 (1988);

X. Campi, {\it Phys.\ Lett.} {\bf B208}, 351 (1988)

\bibitem{lich83}
A.J. Lichtenberg and M.A. Lieberman,
{\it Regular and Stochastic motion}, Springer--Verlag (1983);

Hao Bai--Lin, {\it Chaos},World Scientific (1984);

P. Cvitanovic, {\it Universality in Chaos}, Adam Hilger (1989);

M. Tabor, {\it Chaos and Integrability in Non-linear
Dynamics}, John Wiley (1989);

E. Ott, {\it Chaos in dynamical systems}, Cambridge University
Press (1993);

M. C. Gutzwiller, {\it Chaos in Classical and Quantum Mechanics},
Springer--Verlag (1990);

K. Nakamura, {\it Quantum Chaos -- a new paradigm of non-linear
dynamics},
Cambridge University Press (1993).


\bibitem{bbr1}
G.F. Burgio, M. Baldo and A. Rapisarda,
{\it Phys.\ Lett.\ B} {\bf 321}, 307 (1994)

\bibitem{bbr2}
M. Baldo, G.F. Burgio and A. Rapisarda,
Proceedings of the international conference
{\it Dynamical features of nuclei and finite Fermi systems},
September
13-17,1993, Sitges (Barcelona), Spain, World Scientific (1994).

M. Baldo, G.F.Burgio and A. Rapisarda, in ref.1.

\bibitem{tao94}
M. Baldo, G.F. Burgio and A. Rapisarda,
Proc. of the {\it V International Conference
on Nucleus-Nucleus Collisions}, 30 May-4 June 1994,
Taormina, Italy, and Nucl. Phys. A, in press.

\bibitem{ott93}
E. Ott and T. T\'el, {\it Chaos} {\bf 3}, 417 (1993).

\bibitem{tel87} T. T\'el, {\it Phys. Rev.} {\bf A36}, 1502 (1987).

\bibitem{won}
C. Y. Wong and K. T. R. Davies, {\it Phys. \ Rev. }
{\bf C28}, 240 (1983).

\bibitem{gre}
C. Gregoire {\sl et al.}, {\it Nucl. \ Phys. } {\bf A465},
317 (1987).

\bibitem{ber}
G. F. Bertsch and S. Das Gupta, {\it Phys. \ Rep. }
{\bf 160}, 189 (1988) and refs. therein.

\bibitem{bur88}
G. F. Burgio and M. Di Toro, {\it Nucl.
\ Phys. } {\bf A476}, 189 (1988).


\bibitem{bur92}
G.F. Burgio, Ph. Chomaz and J. Randrup, {\it
Phys.\ Rev.\ Lett.} {\bf 69}, 885 (1992).

\bibitem{ccr}
M. Colonna, Ph. Chomaz and J. Randrup, {\it
Nucl.\ Phys. } {\bf A567}, 637 (1994).


\bibitem{hei93}
H. Heiselberg C.J. Pethick and D.G. Ravenhall,
{\it Phys.\ Rev.\ Lett.} {\bf 61}, 818
(1988) and {\it Ann.\ Phys.} {\bf 223}, 37 (1993).


\bibitem{sau}
G. Sauer, M. Chandra and U. Mosel, {\it Nucl. \ Phys. }
{\bf A264}, 221 (1976)

\bibitem{landau}
M. Baldo, G.F. Burgio and A. Rapisarda, to be published


\bibitem{nota} { It is important to clarify in this context  that
the initial distance between phase space trajectories should
approach to zero. The definition of distance depends on the metric
adopted in phase space, which in our case coincides with eq.(21).
With that definition, the initial distance $d_0$ is very small,
$i.e.$ $d_0 = 10^{-3} fm^{-2}$.}

\bibitem{moo93}
H.T. Moon, {\it Rev.\ Mod.\ Phys.} {\bf 65}, 1535 (1993)


\bibitem{ala94}
See the latest Aladin and EOS data in ref.1.



\end{thebibliography}
\end{document}